\documentclass[12pt,a4paper,final]{iopart}
\usepackage{iopams}
\usepackage{graphicx}
\usepackage[breaklinks=true,colorlinks=true,linkcolor=blue,urlcolor=blue,citecolor=blue]{hyperref}
\usepackage{dcolumn}   
\usepackage{bm}        
\usepackage{amssymb}   
\usepackage{color}
\usepackage{cite}
\usepackage{mcite}

\newcommand{\ria}{\rightarrow}
\newcommand{\ket}[1]{\left\vert #1 \right\rangle}
\newcommand{\bra}[1]{\left\langle #1 \right\vert}

\newcommand{\ee}{\mathrm{e}}
\newcommand{\ii}{\dot{\iota}}
\newcommand{\lam}{\lambda}
\newcommand{\ups}{\upsilon}

\newcommand{\pac}[1]{\left\{ #1 \right\}}
\newcommand{\pap}[1]{\left( #1 \right)}
\newcommand{\pas}[1]{\left[#1 \right]}

\newcommand{\rhoo}{\hat{\rho}}
\newcommand{\Ho}{\hat{H}}
\newcommand{\Jo}{\hat{J}}
\newcommand{\To}{\hat{T}}
\newcommand{\ao}{\hat{a}}
\newcommand{\aod}{\ao^{\dag}}

\begin{document}


\title{Robust quantum correlations in out-of-equilibrium matter-light systems }

\author{O.~L. Acevedo}
\address{Departamento de F\'{i}sica, Universidad de los Andes, A.A. 4976, Bogot\'{a}, Colombia}
\ead{ol.acevedo53@uniandes.edu.co}

\author{L. Quiroga}
\address{Departamento de F\'{i}sica, Universidad de los Andes, A.A. 4976, Bogot\'{a}, Colombia}

\author{F.~J. Rodr\'{i}guez}
\address{Departamento de F\'{i}sica, Universidad de los Andes, A.A. 4976, Bogot\'{a}, Colombia}

\author{N.~F. Johnson}
\address{Department of Physics, University of Miami, Coral Gables, Miami, FL 33124, USA}


\begin{abstract}
High precision macroscopic quantum control in interacting light-matter systems remains a significant goal toward novel information processing and ultra-precise metrology. We show that the out-of-equilibrium behavior of a paradigmatic light-matter system (Dicke model) reveals two successive stages of enhanced quantum correlations beyond the traditional schemes of near-adiabatic and sudden quenches. The first stage features magnification of matter-only and light-only entanglement and squeezing due to effective non-linear self-interactions. The second stage results from a highly entangled light-matter state, with enhanced superradiance and signatures of chaotic and highly quantum states. We show that these new effects scale up consistently with matter system size, and are reliable even in dissipative environments.
\end{abstract}

\maketitle

Many-body quantum dynamics are at the core of many natural and technological phenomena, from understanding of superconductivity or magnetism, to applications in quantum information processing as in adiabatic quantum computing \cite{Dziarmaga}. Critical phenomena, defect formation, symmetry breaking, finite-size scaling are all aspects that emerge from the collective properties of the system \cite{Sachdev}. Spin networks, many-body systems composed of the simplest quantum unit, are an obvious starting point to understand those phenomena, as they enclose much of their complex behavior in a highly controllable and tractable way. However, if the system under investigation includes a radiation subsystem, new opportunities arise for monitoring and characterizing the resulting collective phenomena \cite{WignerBEXP,qubit3000}. By devising driving protocols of the light-matter interaction, high precision macroscopic control then becomes a possibility, regardless of whether the focus is on the matter subsystem, the light, or the composite manipulation of both. This is particularly true for the Dicke model (DM) \cite{Dicke}, which is the subject of the present work.\\

The DM describes a radiation-matter system which, despite its simplicity, exhibits a wide arrange of complex collective phenomena, many of them specifically associated with the existence of a quantum phase transition (QPT) \cite{BrandesPRL,*BrandesPRE}. Experimental realizations of the DM have been presented in different settings, from proposed realizations in circuit quantum electrodynamics \cite{Ciuti,*Marquadt}, to the recent very successful demonstrations of DM superradiance in various cold atom experiments \cite{Baumann,*DMEXPHamner,*DMEXPBaden}. While the light and matter properties in the equilibrium ground state are relatively well known \cite{BrandesENT,ReslenEPL,VidalDicke,Wang2014,OlayaCastroEPL}, its fully quantum out-of-equilibrium critical behavior is just starting to be understood \cite{BastidasDicke,PRLart}. \\

In the DM, both the matter and field are known to act as mediators of an effective non-linear self-interaction involving each other \cite{ReslenEPL,OlayaCastroEPL}. These non-linear interactions produce interesting phenomena in both atomic and optical systems \cite{Kitagawa1993,KerrNAT}. Among the most relevant effects, there is the strong collapse and revival of squeezing \cite{Milburn1,Transverse}, which in many matter states can be related to atom-atom entanglement \cite{Wang}. Applications of such effects are widespread, including high precision quantum metrology \cite{Metrology,*MetrologyOPT,*Rey2007}, and quantum information technologies \cite{QuInCo,*VedralE}. As the non-linear interactions are only effective, it is of essential relevance to understand how the eventual matter-field correlations could affect the generation of the desirable quantum squeezing in each subsystem. We will theoretically address this problem in the particular setting of the finite size DM, when it is ramped across the critical threshold starting from its initial equilibrium non-interacting state.\\

Some previous works have already examined the dynamical emergence of quantum effects on each subsystem of the DM. However they focused on the semi-classical limit and a static coupling after a sudden quench \cite{Altland2,Alvermann,Alvermann1,Furuya,*Song}, small dynamic oscillations around a phase space region \cite{BastidasDicke}, or under time-delayed feedback control \cite{NJPr11,*NJPr12}. Our work goes beyond this by giving a fully quantum analysis of the process of continuously turning on the interaction, in order to assess the emergent effects of the non-linear self interaction in each subsystem. We have carried this out across all dynamical regimes: from the very slow adiabatic regime, where the equilibrium ground state results apply, to the sudden quench regime which is the initial preparation scheme for most previous non-equilibrium results. In between these limits, we have found an unexplored yet very rich intermediate set of annealing velocities featuring remarkable amplification of critical quantum effects as compared to the near-equilibrium results. This magnification of critical properties is followed by a novel chaotic dynamical phase, characterized by giant light-matter entanglement \cite{PRLsub}, negative Wigner quasi-distributions, and signatures of quantum chaos.\\

This paper is structured as follows. Section \ref{secm} presents the model, dynamical setting, and methods of analysis. Section \ref{seca} explains the general profile of the dynamical evolution under the perspective of matter and light quantum properties, namely squeezing and the order parameter (OP). Section \ref{secb} gains further insight from the point of view of phase space representations of the subsystems. Section \ref{secc} establishes the consistency of our results, even under dissipative effects, and for different system sizes, by means of a power-law relation. Section \ref{seconc} concludes with some general remarks.

\section{Theoretical framework}
\label{secm}

The DM describes a totally symmetric interaction of $N$ matter qubits with a radiation mode. Its Hamiltonian is \cite{Dicke},
\begin{equation}
\Ho =\epsilon \Jo_{z}+\omega \ao^{\dag }\ao+ 2 \frac{\lam (t)}{\sqrt{N}} \Jo_{x}\left( \ao ^{\dag }+\ao\right) .
\label{eqHDick}
\end{equation}
Symbols $\Jo_{i}=\frac{1}{2}\sum_{j=1}^{N}\hat{\sigma} _{i}^{\pap{j} }$ represent collective operators of the qubits, and symbol $\ao^{\dag }$ ($\ao$) is the creation (annihilation) operator of the radiation field. Coefficients $\epsilon$ and $\omega$ are single qubit and single mode excitation energies respectively. From now on, we set resonant energies, $\epsilon =\omega =1$. The thermodynamic limit ($N \ria \infty$) phase boundary of the controlled interaction is $\lam_c = \sqrt{\epsilon \omega}/2 = 0.5$, which is slightly different at finite values of $N$ \cite{VidalDicke}. The total system $S$, represented by a density matrix $\rhoo_S$, evolves under a unitary part generated by $\Ho$, and a dissipative part caused by radiation losses of the field cavity. The time evolution is then expressed by a master equation \cite{breuer},
\begin{eqnarray}
\frac{d}{d t}\rhoo_S && = -\ii \pas{\Ho,\rhoo_S} \nonumber \\
&& +\kappa \pap{\bar{n}+1} \pap{2\ao \rhoo_S \aod - \pac{\aod \ao,\rhoo_S}}
+\kappa \bar{n} \pap{2\aod \rhoo_S \ao - \pac{\ao \aod,\rhoo_S}} ,
\label{eqME}
\end{eqnarray}
where $\kappa$ is the damping rate of the cavity, and $\bar{n}$ is the thermal mean photon number. At initial time $t=0$ the system starts at
\begin{equation}
\rhoo (0) = \ket{-N/2}_z\bra{-N/2}_z \otimes \frac{\ee^{-\beta \aod\ao}}{\tr\pac{\ee^{-\beta \aod\ao}}},
\end{equation}
where $\ee^{-\beta} = \bar{n}/\pap{\bar{n}+1}$. Thus, the initial state at  $\lam (0) = 0$ corresponds to an unentangled state formed by the qubit ground state times the field thermal equilibrium state at inverse temperature $\beta=1/(k_B T)$, with $k_B$ the Boltzmann constant. The controlled interaction will change with an annealing velocity (AV) $\ups$ by a linear ramping: $\lam (t) =  \ups t$. The interval of interest is $\lam \in [0,2]$, well across the QPT. This work is focused on subsystem properties. Any subsystem $A$ will be described by a density matrix $\rhoo_A$, defined as the trace with respect to the other degrees of freedom: $ \rhoo_A (t) = \tr _{S-A} \pac{ \rhoo_S }$.\\

We explore the dynamical development of the quantum properties in subsystems of the matter-light system, caused by this ramped QPT crossing, without resorting to any common simplifications like mean-field, rotating wave or semi-classical approximations. Our main results lie at AV values outside the well-known adiabatic and sudden quench regimes, where quantum correlations such as entanglement and squeezing of each separate subsystem can get significantly enhanced. Exact numerical solutions have been obtained by integrating Eq. \ref{eqME}. When $\kappa =0$, the evolution lies on the pure Hilbert space and is generated by the DM Hamiltonian. Otherwise, the evolution lies in the space of density matrices and its generator is the total Liouvillian superoperator of Eq. \ref{eqME} \cite{breuer}. Thus, the dimension of the numerical evolution vector space is squared as soon as the unitary condition is broken. As the dimension of the field Hilbert space is infinite, the Fock basis $\pac{\ket{n}}$ is truncated up to a number where numerical results converge. We have taken advantage of every simplifying condition, such as parity conservation when $\kappa =0$.\\

The state of any set of $M$ qubits is the same (all qubits are equivalent) and lies on the maximal Dicke manifold $\hat{\mathbf{J}}^2=(J+1)J$, with $J=M/2$. The first quantum property of matter that we address is entanglement, which for a 2 qubit subsystem is measured by the Wootters concurrence $c_W$ \cite{Wootters}. The second one is spin squeezing, which is highly related to multipartite entanglement and usually expressed in terms of a squeezing parameter $\xi_q$ \cite{QuaSpiSqu}. Under unitary evolution ($\kappa = 0$), when parity of the total system is even and conserved, $c_W$ and $\xi_q$ are explicitly related whenever the concurrence is different from zero \cite{Wang1}:
 \begin{equation}
1-\xi_q ^2=(N-1) c_W =\frac{2}{N}\left( \left|\left\langle \Jo_{+}^{2}\right\rangle \right|+\left\langle \Jo_{z}^{2}\right\rangle -\frac{N^2}{4} \right).
 \label{eqSqueezequb}
 \end{equation}
The factor $N-1$ in the concurrence is a direct manifestation of the monogamy of entanglement \cite{Monogamy}, since each qubit is equally entangled to any other one and hence the finite amount of possible entanglement is evenly distributed. Any qubit state can be visualized by the Agarwal-Wigner function (AWF), which is a Bloch sphere representation of the qubit's density matrix \cite{WignerQubit1},
 \begin{equation}
 W_q(\theta,\phi)=\sum_{l=0}^{N}\sum_{m=-l}^{l} T_{l,m}Y_{l,m}(\theta,\phi),
 \label{eqWignerqub}
 \end{equation}
and $Y_{l,m}$ are the spherical harmonics. Terms $T_{l,m}=\mathrm{tr}\pac{\rhoo_q \To_{l,m}}$, with $\hat{\rho}_q$ the qubit subsystem density matrix, are the expected values of the multipole operator,
\begin{equation}
\hat{T}_{l,m}=\sum_{M,M^{\prime}=-j}^{j} (-1)^{j-m} \sqrt{2l+1} \pap{\begin{array}{ccc} j & l & j \\ -M  & m & M^{\prime} \end{array}} \ket{jM}\bra{jM^{\prime}},
\end{equation}
where $j=N/2$, and ${\tiny \pap{\begin{array}{ccc}  j & l & j \\ -M  & m & M^{\prime} \end{array}} }$ is the Wigner $3j$ symbol.\\

In addition to the matter subsystems, the DM has the presence of a radiation mode that is usually much more experimentally accessible, thanks to the radiation leaked by the cavity and properties disclosed by tomographic techniques \cite{WignerBEXP}. The general state of the mode can be condensed in the form of its Wigner quasi-probability distribution \cite{quantumoptics,Fuchs},
\begin{equation}
W_b\pap{ \alpha ,\rhoo _b } = \sum_{n=0}^{\infty }\left( -1\right) ^{n}\left\langle
n\right\vert \hat{D}^{\dag }\left( \alpha \right) \rhoo_b \hat{D}\left( \alpha \right)
\left\vert n\right\rangle, \label{eqWignerbos}
\end{equation}
where $\hat{D}\left( \alpha \right) =\mathrm{e}^{\alpha \ao^{\dag }-\alpha ^{\ast }\ao}$ is the displacement operator, and $\alpha \in \mathbb{C}$. There is also an analogous squeezing parameter in the field mode, though it cannot be directly related to a form of entanglement. It is expressed in terms of the variance and covariance of field quadratures \cite{Squeezelight},
\begin{equation}
\xi_b^2 = Var(x)+Var(p)-\sqrt{\left( Var(x)-Var(p)\right)^2+4Cov(x,p)^2}.
\label{eqSqueezebos}
\end{equation}
The displacement operator in Eq. \ref{eqWignerbos} is related to the quadratures by $\sqrt{2}\alpha=x+\ii p$.

\section{Out-of-equilibrium enhancement of quantum correlations}
\label{seca}

\begin{figure}[!h]
  \includegraphics[width=1.00\textwidth]{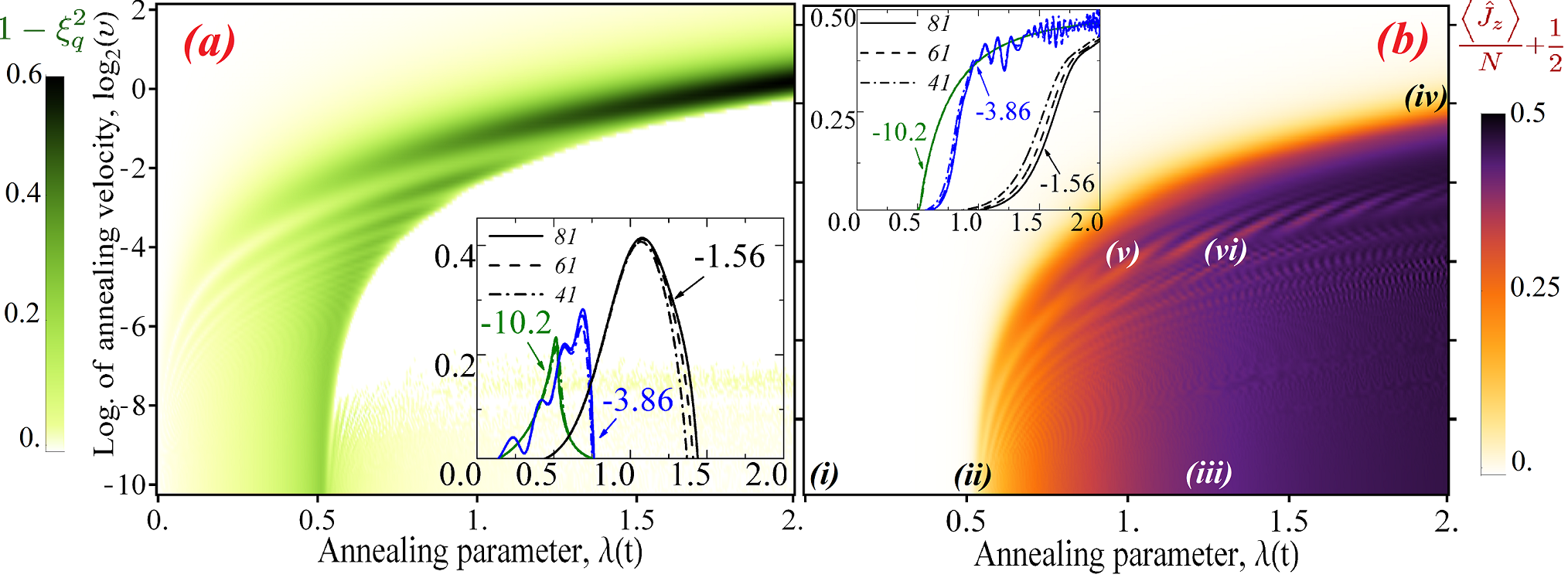}
  \caption{Dynamic profiles (time varies from left to right) of matter subsystem properties. \textbf{(a)} Two qubit concurrence $c_W (N-1)=1-\xi_q^2$, related to spin squeezing in Dicke manifold states (see Eq. \ref{eqSqueezequb}). \textbf{(b)} Scaled order parameter (OP) $ \left\langle \Jo_z \right\rangle/N+1/2 $. Results are for $N=81$, and unitary evolution ($\kappa =0$). Dynamic regimes range from adiabatic (bottom) to sudden quench (top) annealing velocities (AV). Compared to what is possible in near-equilibrium (lower part), dynamic magnification of squeezing occurs at higher AV preceded by the emergence of the order parameter at delayed values of $\lam$. The insets show the evolution for three selected values of $\log_2\upsilon$ (according to color) and three different system sizes (according to line style). Strong similarities exist for different values of $N$, pointing toward scaling properties of the results. Roman symbols in \textbf{(b)} hint at instants where phase space representations of subsystem states are depicted in Fig. \ref{fig3}.}\label{fig1}
  \end{figure}

A general picture of the time evolution of the matter quantum properties is revealed by means of spin squeezing (related to matter entanglement) and the expectation value of its OP in Figs. \ref{fig1}a-b. All the relevant AV are considered in each plot, from the adiabatic limit (bottom of vertical axis) to the sudden quench (top). If the adiabatic condition is fulfilled, the spin squeezing or entanglement of the matter subsystem corresponds to those well known from the equilibrium QPT \cite{Bakeme,BrandesENT}. In the sudden quench limit, there are no appreciable changes since the system essentially stays where it started. In between those limits, a complex dynamical regime emerges. Near the thermodynamic limit equilibrium phase transition $\lambda = 0.5$, adiabatic evolution exhibits a maximum value of entanglement followed by its decay and then the growth of the OP. The anti-correlation between squeezing and OP is present in the dynamical regime as well, so that Figs. \ref{fig1}a and \ref{fig1}b are effectively negative images of each other. As the range of high AV is entered, the position of the maximum point of squeezing and the onset point of the OP are pushed toward higher values of $\lam$. Any value of $\ups$ beyond the adiabatic limit leads to the dynamical regime, provided that the controlled interaction is ramped up to a high enough value. In other words, the sudden quench condition is only a consequence of the upper bound of $\lam$.\\

More important than the position of maximum concurrence in the dynamical regime, is its value. It reaches up to three times the corresponding value of the adiabatic evolution (see Fig. \ref{fig1}a). This is a remarkable improvement for the collective generation of such a distinctive quantum property, and is due entirely to the system being in the non-equilibrium dynamical regime. For some given AV values, this magnified concurrence for different number of qubits $N$ follows closely the scaling $c_W (N-1)$ (see inset of Fig. 1a), keeping the spin squeezing parameter curves almost size independent, and gets modestly better as the system size grows. Some size dependent deviations are present in curves for the $\log_2 \ups  = -1.56$ case. The sudden death point of concurrence is also virtually size-independent. The scaling properties of the dynamical regime will be discussed in section \ref{secc}. The matter OP, for some chosen AVs, also shows very good scaling collapse in its dynamical evolution, i.e. the curves in the inset of Fig. \ref{fig1}b are essentially size independent thanks to the scaling $\Jo_z/N$ in the OP.\\

With respect to the maximum matter OP in Fig. \ref{fig1}b, it is bounded. Therefore independent of the regime, it cannot be increased. The kind of enhancement revealed by the OP is a magnification of \emph{superradiance}, which is the traditional way to describe the QPT of DM \cite{lieb}. The OP indicates the scaled number of excitations within the subsystem, which is bounded because of finiteness of $N$. In the dynamical regime, excitations develop more quickly (see blue curve in inset of Fig. \ref{fig1}b). In the case of eventual leakage of excitations through the cavity (see Section 4 for cavity loss effects), it manifests itself as a sharp burst of photons which are suddenly released. Bursts like these are excellent indicators of the DM QPT in cold atom experimental realizations \cite{Baumann}.\\

\begin{figure}[!h]
  \includegraphics[width=\textwidth]{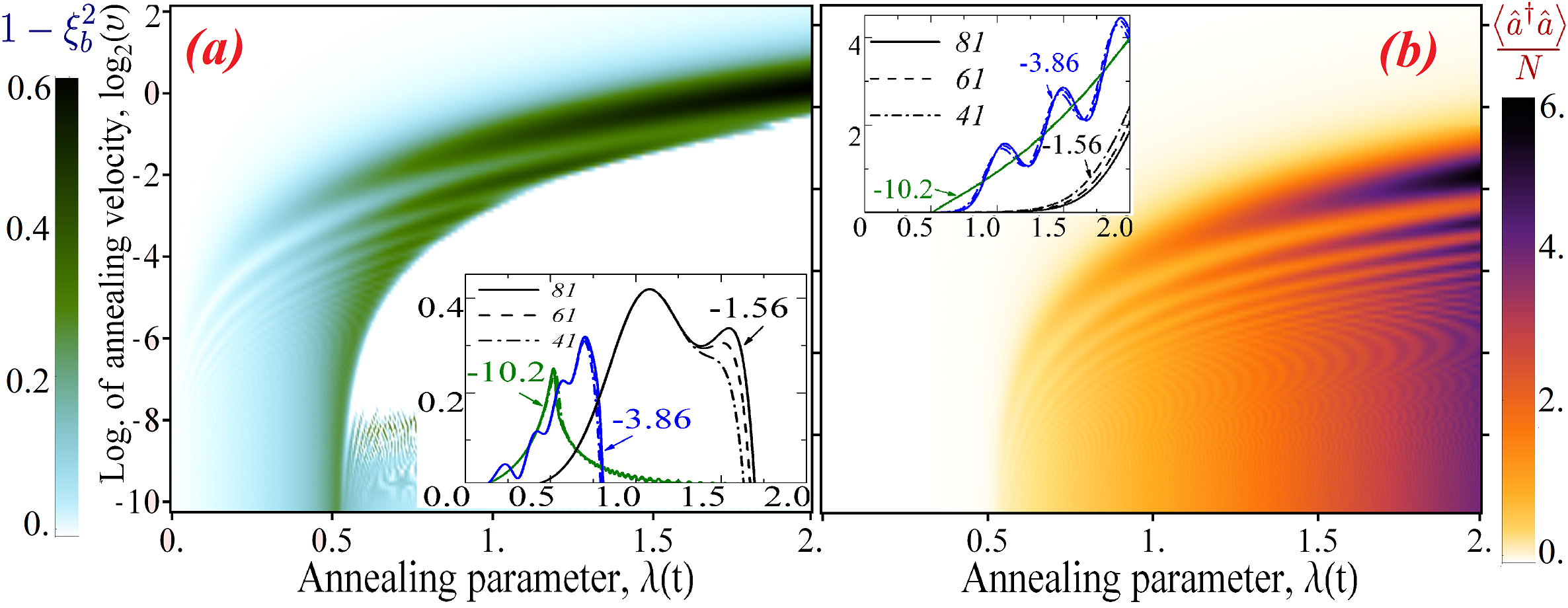}
  \caption{Corresponding dynamic profiles and insets of Figs. \ref{fig1}a and \ref{fig2}b, now for the field subsystem. \textbf{(a)} Evolution of $1-\xi_b^2$, as defined in Eq. \ref{eqSqueezebos}, whenever it is greater than zero (squeezed radiation). \textbf{(b)} Scaled OP of light $\left\langle \ao^{\dag}\ao \right\rangle /N$. Major tendencies of Fig. \ref{fig1} are replicated here. However the unbounded nature of the field OP allows an appreciation of the field intensity's oscillation (associated with supperradiance) around the equilibrium values, with intervals of significant dynamical enhancement of superradiant bursts. The oscillatory behavior is related to a combined light-matter quantum chaotic stage.} \label{fig2}
  \end{figure}

Qubit quantum properties are very important for applications in quantum computation and quantum information. However, the finite character of the matter subsystem conceals much of the complexity of the evolution. The field subsystem, with its own relevance in highly controlled quantum optics, does not have this restriction. Field dynamical profiles analogous to that of Fig. \ref{fig1} are shown in Figs. \ref{fig2}a-b. Similarities with Figs. \ref{fig1}a-b are noticeable, though with relevant differences. A large dynamical magnification of the field quadrature squeezing is also present, but it dies off later than its matter counterpart. Thus the development of the field OP is delayed as compared to that of the qubits. Superradiance, as seen by value of the field OP, is now not only sharper, but it can be as much as twice the value attainable with adiabatic ramping at a given $\lam$. This greatly enhances the intensity of superradiant bursts. Radiation intensity enhancements of this kind have also been predicted in the DM submitted to sudden quenches and a lossy cavity \cite{Fuchs}. Dynamically enhanced and suppressed superradiance alternate after the first burst (see blue curve in inset of Fig. \ref{fig2}b), a behavior absent in the equilibrium QPT. A wavy plateau in figure \ref{fig1}b coincides with that alternating stage. The dynamical phase marked by these oscillations is characterized by giant light-matter entanglement, with emergent quantum properties beyond the critical ones \cite{PRLsub}.\\

Magnification of squeezing and the other effects discussed so far are caused by a prolonged realization of effective non-linear interactions within both subsystems, after a preparation stage where the sudden quench approximation holds. Then, the non-linear processes are realized as one axis squeezing with a transverse field for matter \cite{QuaSpiSqu,Transverse}, and Kerr-like interactions for radiation \cite{KerrNAT}. During these separate squeezing processes, both radiation and matter act as effective interaction mediators for each other. This extends to the non-equilibrium case static correspondences with matter-only \cite{ReslenEPL}, or radiation-only systems \cite{OlayaCastroEPL}. Separate squeezing, which is related to internal quantum entanglement, suddenly dies as soon as combined matter-light entanglement emerges, because of monogamy and the breaking of the effective non-linear interaction condition. As the field has more information capacity, its squeezing can survive longer. By contrast, in the adiabatic regime, entanglement within and between subsystems compete against each other, because they occur simultaneously at the critical point.\\

\section{Matter-light quasi-probability behavior}
\label{secb}

\begin{figure}[!h]
  \includegraphics[width=0.8 \textwidth]{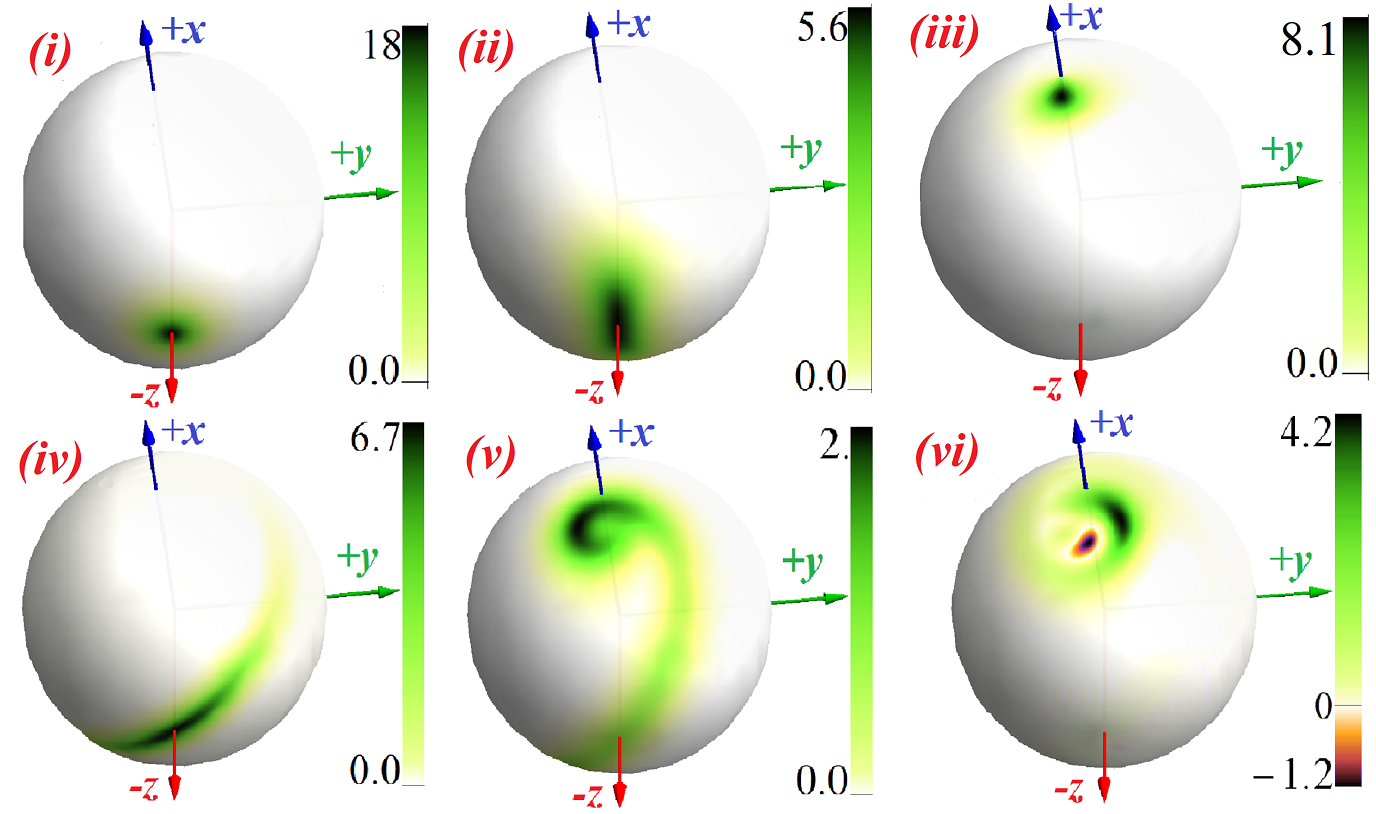}
  \caption{Agarwal-Wigner $W_q$ at the instants indicated in Fig. \ref{fig1}b by the corresponding number (dissipative effects are ignored, $\kappa=0$). They are phase space representations of the matter subsystem, depicted on the Bloch sphere. Color scale of all Wigner functions can be very different. Maximal dynamical squeezing in \textbf{(iv)} is much greater than the adiabatic counterpart in \textbf{(ii)}. Also, negative complex structures develop in the combined light-matter chaotic stage in \textbf{(v-vi)}; which contrast with an adiabatic ordered state as in \textbf{(iii)}. The stage in \textbf{(iii,v,vi)} has also no squeezing and no concurrence, since the Wigner function is now split along the $x$ and $-x$ directions and no longer concentrated around the initial state. Opposite Bloch hemispheres are not shown because of symmetry: $W_q(\theta,\phi+\pi)=W_q(\theta,\phi)$.}\label{fig3}
  \end{figure}

\begin{figure}[!h]
  \includegraphics[width=\textwidth]{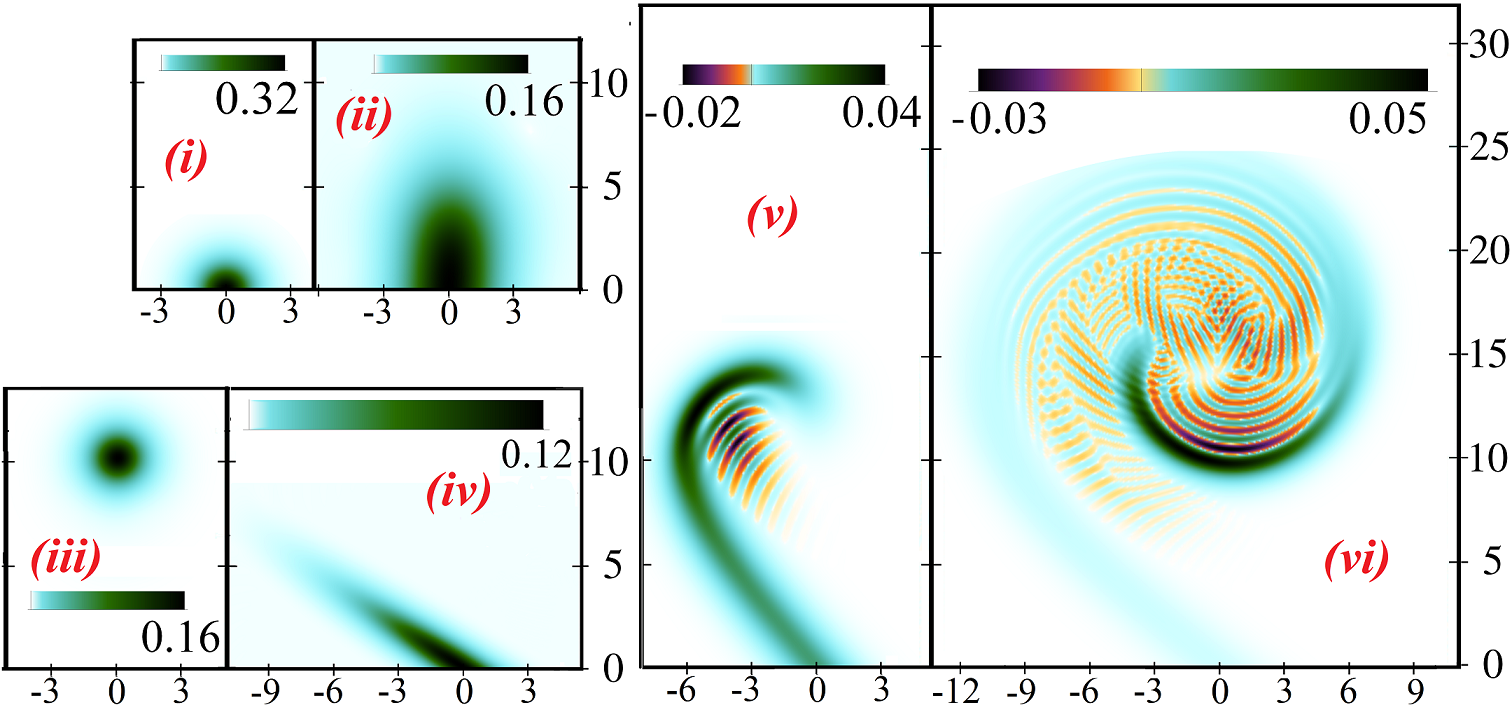}
  \caption{Analogous pictures of Fig. \ref{fig3} but for the field Wigner distribution $W_b$. This is represented in the $x-p$ plane of position (vertical) and momentum (horizontal) quadrature. All figures have the same $x-p$ scale but color scale can be very different. The chaotic stage in \textbf{(iv-v)} exhibits sub-Planck structures. Negative regions of $x$ are not shown because of symmetry: $W_b(x,p)=W_b(-x,-p)$.}\label{fig4}
  \end{figure}

Complete state representations of the subsystems state reveal more details of the processes involved in the dynamical regime. Figures \ref{fig3} and \ref{fig4} show several snapshots of Wigner and AWF quasi-distributions at different instants. The enormous difference in the squeezing amount between the dynamical regime and the adiabatic one, as well as a difference in the direction of squeezing, is now graphically clear. This can be seen by comparing Figs. \ref{fig3}-\ref{fig4} (iv)  with Figs. \ref{fig3}-\ref{fig4} (ii), and was previously shown for different AV values in Figs.\ref{fig1} and \ref{fig2}. In any regime, death of squeezing is associated with a splitting in half of the distributions, caused by the spontaneous symmetry breaking of the QPT \cite{PRLsub}. This splitting is symmetrical along the $x$ and $-x$ direction in both subsystems. The system's density of excitations, indicated by the OP, increases as the distributions become displaced away from the initial state.\\

Features of the complex light-matter entangled stage are clarified by Figs. \ref{fig3}-\ref{fig4} (v-vi), which contrast with the adiabatic splitting in Figs. \ref{fig3}-\ref{fig4} (iii). The dynamical splitting of the distribution leads to negative scars and complex patterns for both subsystems, though the phenomenon is far more conspicuous in radiation. Donut shapes with a negative centered AWF such as that in Fig. \ref{fig3} (vi), have been experimentally obtained in 3000 atoms with just a single photon \cite{qubit3000}. In addition, round-tailed interference patterns such as that in Fig. \ref{fig4} (vi) have been obtained for light in Kerr-like media following a Fokker-Planck equation \cite{Milburn1}, which confirms the presence of non-linear effective interactions. These results mean that the field is able to exhibit chaotic behavior itself, regardless of its entanglement to the matter subsystem. Despite being a single radiation mode, the field acts as a reservoir that dissipates the quantum correlations present in the squeezed states of each subsystem. If the qubits were not coupled to the field, squeezing could have revivals after its sudden death \cite{Transverse}. The field Wigner function is not only full of negatives regions (a marker of non-classical light), but it also contains abundant so-called sub-Planckian structures which have been related to quantum chaos \cite{ZurekSUBP}. The finite-size DM is non-integrable, and its ordered phase has been connected to chaotic features \cite{BrandesPRL,Qin2014}. This chaotic onset is responsible for the small size dependence on curves in insets of Figs. \ref{fig1} and \ref{fig2} for the $\log_2 \ups  = -1.56$ case, since the symmetry breaking occurs deeper in the ordered phase. As spin squeezing is very sensitive to chaos \cite{Furuya,*Song}, it is no surprise that its sudden death becomes irreversible once the light-matter entangled stage arises.\\

\section{Scaling and robustness of non-equilibrium quantum correlations}
\label{secc}

\subsection{Robustness of results}

\begin{figure}[!h]
  \includegraphics[width=\textwidth]{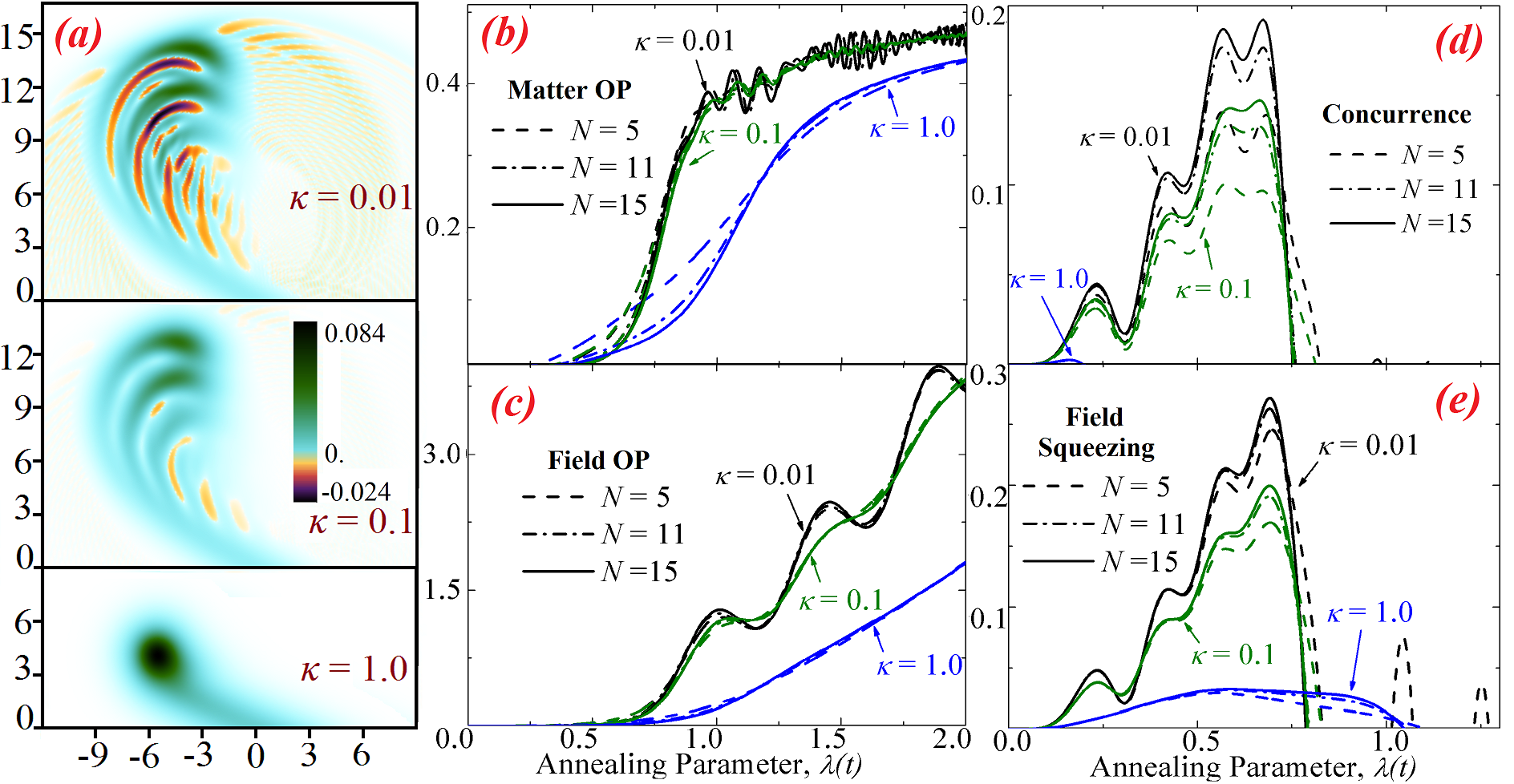}
  \caption{Effect of a leaky field cavity (expressed in terms of cavity decay rate $\kappa$) on the dynamical enhancement of quantum properties. \textbf{(a)} Wigner distribution for AV $\log_2 \upsilon = -1.58$, $N=15$ qubits at instant $\lam = 2$. \textbf{(b-e)} Different field and matter properties as function of time for $\log_2 \upsilon =-3.86$, and three different system sizes. Quantum chaos signatures (sub-Planckian structures, negative Wigner distributions, and oscillations in OPs) are highly sensitive to dissipative perturbations. On the other hand, squeezing and internal entanglement effects are far more robust, especially for field squeezing. In all figures $\bar{n}=0$.} \label{fig5}
  \end{figure}

So far, unitary evolution have been assumed in this work. Figures \ref{fig5}a-e address how the presence of losses in the cavity affect the dynamics. All the main results found in $\kappa=0$ evolution survive very well if decoherence towards the environment is two order of magnitudes lower than the main energy scale. Even if dissipation is at values of just an order of magnitude below, spin squeezing effects remain highly robust, with increasing noise resistance with system size. Field squeezing surprisingly survives to dissipation regimes comparable to the Hamiltonian dynamics itself. On the other hand, detailed features of the chaotic stage (such as OP oscillations, negative regions, and sub-Planckian structures) are far more sensitive to decoherence, requiring losses to be at levels of $\kappa = 0.01$. These very sensitive features have been proposed as tools for measuring very weak forces \cite{Toscano}. In our analysis, we have found that small finite values of $\bar{n}$ (such as those typical at the ultra-low temperatures of most experimental realizations) do not change dramatically the conclusions; they just slightly intensify the process of decoherence. Previous works on non-equilibrium DM under different settings have also confirmed that some major phenomena present here, such as spin squeezing, survive losses. Enhanced superradiance can be detected \cite{Fuchs}, and squeezing can still have time to develop \cite{Alvermann,Milburn1}. We note that, despite perturbing the quantum state of the system, a lossy cavity compensates by providing the experimental possibility to monitor the dynamic evolution of the field in time \cite{WignerBEXP}.\\

The robustness of quantum correlations suggests that some of the effects described in this paper may be accessible under current experimental realizations of the DM. In particular, Klinder et al. have proven that the kind of control described here as $\lam (t)$ can be realized as a pump laser power increasing over time, and that the noise present in current experiments (nearly analogous to our $\kappa = 0.1$ case) is small enough that the squeezing related quantum correlations predicted in our work could be experimentally assessed \cite{PNAS}. As field squeezing is the most robust quantum correlation (see Fig. \ref{fig5}e) and it is experimentally accessible thanks to a leaky mirror, tomographic techniques and photon statistics could be the first front for experimental comparisons with our predictions \cite{WignerBEXP}. Other tomographic techniques could also analyze properties of the matter subsystem state \cite{WignerQEXP}. Given the specific degrees of freedom forming the qubit states in current realizations (basically matter wave modes) \cite{PNAS}, suitable equivalences of those state resolving techniques to this kind of degrees of freedom should be devised.

\subsection{Scaling of results}

\begin{figure}[!h]
  \includegraphics[width=\textwidth]{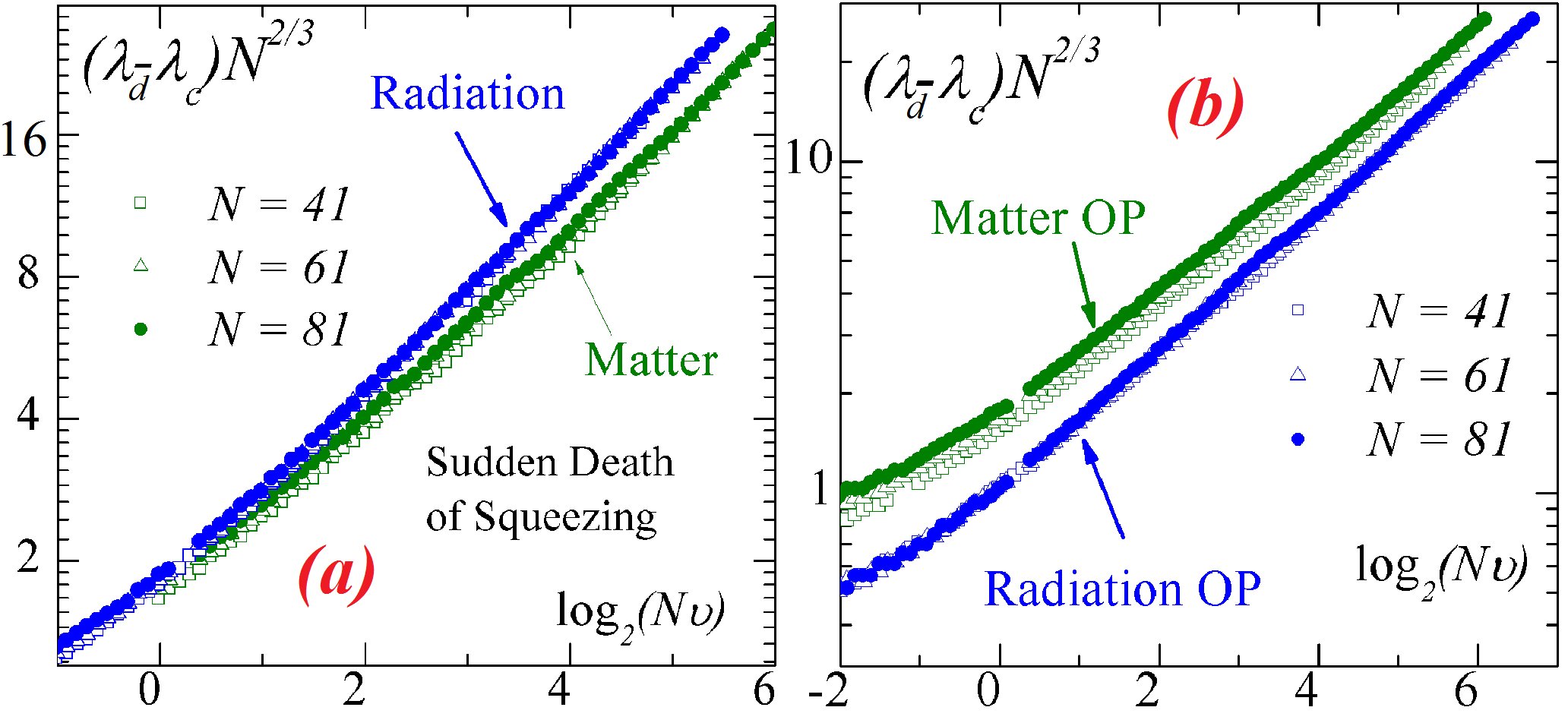}
  \caption{Size-independence of the instant $\lambda_d$ when field and qubits organize in a combined light-matter excited state. \textbf{(a)} Instant of a sudden death of squeezing, in both matter and radiation subsystems. \textbf{(b)} Instant when each system reaches a significant value of its OP. These values are $\left\langle \Jo_z \right\rangle/N+1/2= 0.1$ and $\left\langle \ao^{\dag}\ao \right\rangle / N=0.0123$.  At sufficiently high AVs, all the instances chosen are related by the same scaling power law relation $(\lam_d - \lam_c) \propto \upsilon^{2/3}$.} \label{fig6}
  \end{figure}

Our results exhibit substantial size invariance, as the insets of Figs. \ref{fig1} and \ref{fig2} show. Even the small system sizes in lossy cavity results in Figs. \ref{fig5}b-e are highly size independent, which justifies our claim that the conclusions drawn there can be taken as general trends. Figures \ref{fig6}a-b  show that the onset point of the dynamical regime occurs at points  $\lam_d$  following a power law, $(\lam_d - \lam_c) \propto \upsilon^{2/3}$. This kind of scaling relation is typical in QPT critical properties \cite{Sachdev}. This scaling laws extends our previous results on adiabatic scaling theory \cite{PRLart} to the fast regime. Even though we have used the same kind of scaling variables as in the near adiabatic case, many results in the dynamic regime collapse very well without any re-scaling of $\lam$.\\

The origin of the scaling exponent can be understood in a rather simple way, using concepts from the Kibble-Zurek mechanism \cite{Dziarmaga}. The scaled time is the instant where the healing time of the system (as measured by the inverse mean energy gap in the spectrum $\delta^{-1}$) has the same order of magnitude, i.e. when $t\delta \approx 1$. As $\delta \propto (\lam_d-\lam_c)^{1/2}$ in the ordered phase \cite{ReslenEPL}, and $t \propto (\lam_d-\lam_c)\upsilon^{-1}$, we get the exponent $2/3$ present in all lines of Figs. \ref{fig6}a-b. The fact that equilibrium critical exponents have been used reveals that, even though the emergence of the dynamical regime is well inside the $\lam > 0.5$ range, its mechanism is still related to the QPT, and so many critical phenomena get dynamically enhanced. On the other hand, the fact that the evolutions are very similar in the chaotic region, despite the different sizes, has to do with the relatively equal structure of the interaction well inside the superradiant phase, provided it has not bounced against the finite limits of the matter subsystem.

\section{Conclusions}
\label{seconc}

We investigated the details of the microscopic properties of both matter and light subsystems of the DM when macroscopically driven into the superradiant phase from an initial non-interacting equilibrium state. We have found that, in between the traditional driving schemes (either near-adiabatic or sudden quench), a previously unnoticed intermediate annealing regime emerges, and that this regime can be seen as advantageous for many quantum control processes. The onset of this intermediate phase is marked by induced non-linear self-interactions in both matter and field subsystems, developing squeezed states in each of them that are related to internal entanglement. This squeezing process is much stronger than what can be achieved in near-adiabatic evolutions, since the internal entanglement does not have to compete against the matter-field entanglement that arises in later stages.\\

We have also found that the dynamical squeezing mechanism is succeeded by the development of a chaotic and entangled radiation-matter dynamical phase. Clear signatures of quantum chaos such as fragmented Wigner functions, have been identified in this phase. This stage has its own dynamically enhanced properties such as superradiance, phase order, and sensibility to weak forces. We have established the robustness of all the quantum enhancing processes under dispersive conditions and its invariance to system size. We have shown that this last invariance condenses into a power-law relation associated with critical exponents. We hope this understanding will prove important for developing schemes to generate squeezed and other entangled states, as have been proposed in the contexts of quantum metrology and quantum information processing.

\section{Acknowledgements}

O.L.A, L.Q. and F.J.R. acknowledge financial support from Proyectos Semilla-Facultad de Ciencias at
Universidad de los Andes (2010-2014) and project {\it Quantum control of non-equilibrium hybrid
systems}, UniAndes-2013. O.~L.~A. acknowledges financial support from Colciencias, Convocatoria 511.

\section*{References}
\bibliography{BIBArtSubSistemas}

\providecommand{\newblock}{}
\begin{thebibliography}{10}
\expandafter\ifx\csname url\endcsname\relax
  \def\url#1{{\tt #1}}\fi
\expandafter\ifx\csname urlprefix\endcsname\relax\def\urlprefix{URL }\fi
\providecommand{\eprint}[2][]{\url{#2}}

\bibitem{Dziarmaga}
Dziarmaga J 2010 {\em Adv. Phys.\/} {\bf 59} 1063--1189

\bibitem{Sachdev}
Sachdev S 2011 {\em Quantum Phase Transitions\/} (Cambridge University Press)

\bibitem{WignerBEXP}
Lvovsky A~I 2009 {\em Rev. Mod. Phys.\/} {\bf 81} 299--332

\bibitem{qubit3000}
McConnell R, Zhang H, Hu J, \'{C}uk S and Vuleti\'{c} V 2015 {\em Nature\/}
  {\bf 519} 439--442

\bibitem{Dicke}
Dicke R~H 1954 {\em Phys. Rev.\/} {\bf 93} 99--110

\bibitem{BrandesPRL}
Emary C and Brandes T 2003 {\em Phys. Rev. Lett.\/} {\bf 90} 044101

\bibitem{BrandesPRE}
Emary C and Brandes T 2003 {\em Phys. Rev. E\/} {\bf 67} 066203

\bibitem{Ciuti}
Nataf P and Ciuti C 2010 {\em Nature Comm.\/} {\bf 1} 1

\bibitem{Marquadt}
Viehmann O, von Delft J and Marquardt F 2011 {\em Phys. Rev. Lett.\/} {\bf 107}
  113602

\bibitem{Baumann}
Baumann K, Guerlin C, Brennecke F and Esslinger T 2010 {\em Nature\/} {\bf 464}
  1301--1306

\bibitem{DMEXPHamner}
Hamner C, Qu C, Zhang Y, Chang J, Gong M, Zhang C and Engels P 2014 {\em Nature
  Comm.\/} {\bf 5} 4023

\bibitem{DMEXPBaden}
Baden M~P, Arnold K~J, Grimsmo A~L, Parkins S and Barrett M~D 2014 {\em Phys.
  Rev. Lett.\/} {\bf 113} 020408

\bibitem{BrandesENT}
Lambert N, Emary C and Brandes T 2004 {\em Phys. Rev. Lett.\/} {\bf 92} 073602

\bibitem{ReslenEPL}
Reslen J, Quiroga L and Johnson N~F 2005 {\em Europhys. Lett.\/} {\bf 69} 8

\bibitem{VidalDicke}
Vidal J and Dusuel S 2006 {\em Europhys. Lett.\/} {\bf 74} 817

\bibitem{Wang2014}
Wang T~L, Wu L~N, Yang W, Jin G~R, Lambert N and Nori F 2014 {\em New J.
  Phys.\/} {\bf 16} 063039

\bibitem{OlayaCastroEPL}
Jarrett T~C, Olaya-Castro A and Johnson N~F 2007 {\em Europhys. Lett.\/} {\bf
  77} 34001

\bibitem{BastidasDicke}
Bastidas V~M, Emary C, Regler B and Brandes T 2012 {\em Phys. Rev. Lett.\/}
  {\bf 108} 043003

\bibitem{PRLart}
Acevedo O~L, Quiroga L, Rodr\'{i}guez F~J and Johnson N~F 2014 {\em Phys. Rev.
  Lett.\/} {\bf 112} 030403

\bibitem{Kitagawa1993}
Kitagawa M and Ueda M 1993 {\em Phys. Rev. A\/} {\bf 47} 5138

\bibitem{KerrNAT}
Kirchmair G, Vlastakis B, Leghtas Z, Nigg S~E, Paik H, Ginossar E, Mirrahimi M,
  Frunzio L, Girvin S~M and Schoelkopf R~J 2013 {\em Nature\/} {\bf 495}
  205--09

\bibitem{Milburn1}
Stobi\'{n}ska M, Milburn G and W\'{o}dkiewicz K 2008 {\em Phys. Rev. A\/} {\bf
  78} 013810

\bibitem{Transverse}
Law C, Ng H and Leung P 2001 {\em Phys. Rev. A\/} {\bf 63} 055601

\bibitem{Wang}
Wang X and M{\o}lmer K 2002 {\em Europ. Phys. J. D\/} {\bf 18} 385--391

\bibitem{Metrology}
Giovannetti V, Lloyd S and Maccone L 2011 {\em Nature Photon.\/} {\bf 5}
  222--229

\bibitem{MetrologyOPT}
Schnabel R, Mavalvala N, McClelland D~E and Lam P~K 2010 {\em Nature Comm.\/}
  {\bf 1} 121

\bibitem{Rey2007}
Rey A~M, Jiang L and Lukin M 2007 {\em Phys. Rev. A\/} {\bf 76} 053617

\bibitem{QuInCo}
Braunstein S~L 2005 {\em Rev. Mod. Phys.\/} {\bf 77} 513--577

\bibitem{VedralE}
Amico L, Osterloh A and Vedral V 2008 {\em Rev. Mod. Phys.\/} {\bf 80} 517--576

\bibitem{Altland2}
Altland A and Haake F 2012 {\em New J. Phys.\/} {\bf 14} 073011

\bibitem{Alvermann}
Alvermann A, Bakemeier L and Fehske H 2012 {\em Phys. Rev. A\/} {\bf 85} 043803

\bibitem{Alvermann1}
Bakemeier L, Alvermann A and Fehske H 2013 {\em Phys. Rev. A\/} {\bf 88} 043835

\bibitem{Furuya}
Furuya K, Nemes M and Pellegrino G 1998 {\em Phys. Rev. Lett.\/} {\bf 80} 5524

\bibitem{Song}
Song L, Yan D, Ma J and Wang X 2009 {\em Phys. Rev. E\/} {\bf 79} 046220

\bibitem{NJPr11}
Grimsmo A~L, Parkins A~S and Skagerstam B~S 2014 {\em New J. Phys.\/} {\bf 16}
  065004

\bibitem{NJPr12}
Kopylov W, Emary C, Sch\"{o}ll E and Brandes T 2015 {\em New J. Phys.\/} {\bf
  17} 013040

\bibitem{PRLsub}
Acevedo O~L, Quiroga L, Rodr\'{i}guez F~J and Johnson N~F 2015
  (\textit{Preprint} \eprint{arXiv:1503.05470 [quant-ph]})

\bibitem{breuer}
Breuer H~P and Petruccione F 2007 {\em The Theory of Open Quantum Systems\/}
  (Oxford University Press)

\bibitem{Wootters}
Wootters W~K 1998 {\em Phys. Rev. Lett.\/} {\bf 80} 2245

\bibitem{QuaSpiSqu}
Ma J, Wang X, Sun C and Nori F 2011 {\em Phys. Rep.\/} {\bf 509} 89--165

\bibitem{Wang1}
Wang X and Sanders B 2003 {\em Phys. Rev. A\/} {\bf 68} 012101

\bibitem{Monogamy}
Koashi M and Winter A 2004 {\em Phys. Rev. A\/} {\bf 69} 022309

\bibitem{WignerQubit1}
Dowling J, Agarwal G and Schleich W 1994 {\em Phys. Rev. A\/} {\bf 49} 4101

\bibitem{quantumoptics}
Walls D and Milburn G~J 2008 {\em Quantum Optics\/} (Springer)

\bibitem{Fuchs}
Fuchs S, Ankerhold J, Blencowe M and Kubala B 2015  (\textit{Preprint}
  \eprint{arXiv:1501.07841 [quant-ph]})

\bibitem{Squeezelight}
Walls D~F 1983 {\em Nature\/} {\bf 306} 141--146

\bibitem{Bakeme}
Bakemeier L, Alvermann A and Fehske H 2012 {\em Phys. Rev. A\/} {\bf 85} 043821

\bibitem{lieb}
Hepp K and Lieb E~H 1973 {\em Ann. Phys.\/} {\bf 76} 360--404

\bibitem{ZurekSUBP}
Zurek W~H 2001 {\em Nature\/} {\bf 412} 712--717

\bibitem{Qin2014}
Qin P, Wang W~g, Benenti G and Casati G 2014 {\em Phys. Rev. E\/} {\bf 89}
  032120

\bibitem{Toscano}
Toscano F, Dalvit D, Davidovich L and Zurek W 2006 {\em Phys. Rev. A\/} {\bf
  73} 023803

\bibitem{PNAS}
Klinder J, Ke{\ss}ler H, Wolke M, Mathey L and Hemmerich A 2015 {\em Proc.
  Natl. Acad. Sci. U. S. A.\/} {\bf 112}(11) 3290--3295

\bibitem{WignerQEXP}
Schmied R and Treutlein P 2011 {\em New J. Phys.\/} {\bf 13} 065019

\end{thebibliography}

\end{document}